\documentclass[fleqn,usenatbib]{mnras}

\usepackage[normalem]{ulem}

\usepackage[T1]{fontenc}
\usepackage{ae,aecompl}

\usepackage{graphicx}	
\usepackage{amssymb}	
\usepackage{hyperref}
\hypersetup{
  colorlinks   = true, 
  urlcolor     = blue, 
  linkcolor    = blue, 
  citecolor   = blue 
}
\usepackage{color}

\newcommand\ion[2]{#1$ ${\scshape{#2}}}
\newcommand{\kms}{\,km\,s$^{-1}$}	
\newcommand{\caps}[1]{{\scshape{#1}}}
\def\msun{M$_{\odot}$}

\def\apj{ApJ}
\def\aj{AJ}
\def\mnras{MNRAS}

\def\aap{A\&A}
\def\apjs{ApJS}
\def\anstat{Ann. Statist.}               

\title[Doppler Tomography and Photometry of 1RXS J064434.5+334451]{Doppler Tomography and Photometry of the\\ Cataclysmic Variable 1RXS J064434.5+334451}
\author[J. V. Hern\'andez Santisteban et al.]{J. V. Hern\'andez Santisteban$^{1,2}$\thanks{E-mail:
j.v.hernandez@soton.ac.uk (JVHS); jer@astro.unam.mx (JE); rmm@astrosen.unam.mx (RM); costero@astro.unam.mx (RC)}, J. Echevarr\'ia$^{2\star}$, R. Michel$^{3\star}$ and R. Costero$^{2\star}$\\
$^{1}$Department of Physics \& Astronomy, University of Southampton, Southampton SO17 1BJ, UK\\
$^{2}$Instituto de Astronom\'ia, Universidad Nacional Aut\'onoma de M\'exico, Apartado Postal 70-264, \\Ciudad Universitaria, M\'exico D.F., C.P. 04510, M\'exico.\\
$^{3}$Instituto de Astronom\'ia, Universidad Nacional Aut\'onoma de M\'exico, Apartado Postal 877, \\Ensenada, Baja California, C.P. 22830 M\'exico.}

\begin{document}

\date{Submitted: \today}

\pagerange{\pageref{firstpage}--\pageref{lastpage}} \pubyear{2016}

\maketitle

\label{firstpage}

\begin{abstract}
We have obtained simultaneous photometric and spectroscopic observations of the cataclysmic variable 1RXS J064434.5+334451. We have calibrated the spectra for slit losses using the simultaneous photometry allowing to construct reliable Doppler images from  H$\alpha$ and \ion{He}{ii} 4686 \AA\@ emission lines. We have improved the ephemeris of the object based on new photometric eclipse timings, obtaining  $HJD = 2453403.759533 + 0.26937446E$. Some eclipses present a clear internal structure which we attribute to a central \ion{He}{ii} emission region surrounding the white dwarf, a finding supported by the Doppler tomography. This indicates that the system has a large inclination angle $i=78 \pm 2^{\circ}$. We have also analyzed the radial velocity curve from the emission lines to measure its semi--amplitude, $K_1$, from H$\alpha$ and \ion{He}{ii} 4686 and derive the masses of the components: $M_1=0.82\pm0.06$~M$_{\odot}$, $M_2=0.78\pm0.04$ M$_{\odot}$ and their separation $a=2.01\pm0.06$ $R_{\odot}$. The Doppler tomography and other observed features in this nova-like system strongly suggests that this is a SW Sex-type system.
\end{abstract}

\begin{keywords}
accretion -- method: Doppler tomography -- stars: cataclysmic variables, J0644+3344, SW Sex-type stars
\end{keywords}

\section{Introduction}
Cataclysmic variables (CV) are semi-detached binary systems which consists of a white dwarf (WD) primary surrounded by a Keplerian disc accreted from a Roche Lobe-filling late-type secondary star. In systems where the mass transfer is high \citep[$\dot{M}\sim10^{-8}$~M$_{\odot}$~yr$^{-1}$,][]{tow09}, the disc will become steady and remain bright for longer periods, suppressing the typical outbursts of CVs with lower mass transfer. These systems are commonly referred as \textit{nova-like variables} (NL). \citet{tho91} constructed an initial qualitative description of NL systems which possess V-shaped eclipses, single-peaked lines and lags between the photometric and spectroscopic ephemeris, known as SW Sex stars. Recently, this list of properties had been revised in order to account for an increasing variety of systems (including non-eclipsing) that share similar traits\footnote{See D.~W.\ Hoard's Big List of SW Sextantis Stars at \url{http://www.dwhoard.com/biglist}.}. Furthermore, SW Sex stars seem to be the dominant population of systems with orbital periods around 3-4 hr \citep{rod07} and their possible connection to nova eruptions may provide information in the general context of CV evolution \citep{pat13}.

1RXS J064434.5+334451 (hereinafter J0644) is a bright object ($V\sim$13.3), discovered during the Northern Sky Variability Survey  by \citet{wea04} (NSVS 7178256), was initially identified as a $\beta$ Lyrae object by \citet{hea08}, but they point out that \citet{sea07} (hereinafter S07)  have identified the object as a deep-eclipsing CV with an orbital period of nearly 6.5 hr.  The latter authors present the first spectroscopic and photometric study of this object and derive radial velocity semi-amplitudes for the primary and secondary stars, from which they obtained a mass ratio $q=0.78$, and individual stellar masses in the range 0.63-0.69 $M_{\odot}$ for the white dwarf and 0.49-0.54 $M_{\odot}$ for the late--type star. They found that this nova--like resembles a UX UMa or a SW Sex type object.

In this paper, we present new spectroscopic and simultaneous photometry of J0644. We revisit the ephemeris of the system based on new observed eclipses, whose shapes are discussed in detail. We discuss the radial velocity measurements of the Balmer H$_{\alpha}$ and the high-excitation \ion{He}{ii} $\lambda 4686$ emission line. We present Doppler tomography reconstructions, calibrated with the simultaneous photometry. Finally, a discussion is made on the classification of the object among the CVs, which point out towards the group of the SW Sex--type stars.

\section{Observations and Reduction}
\label{sec:observations}
 
\subsection{Spectroscopy}
\begin{figure*}
\begin{center}
\includegraphics[trim=0cm 0.5cm 0cm 0cm, clip, width=16cm]{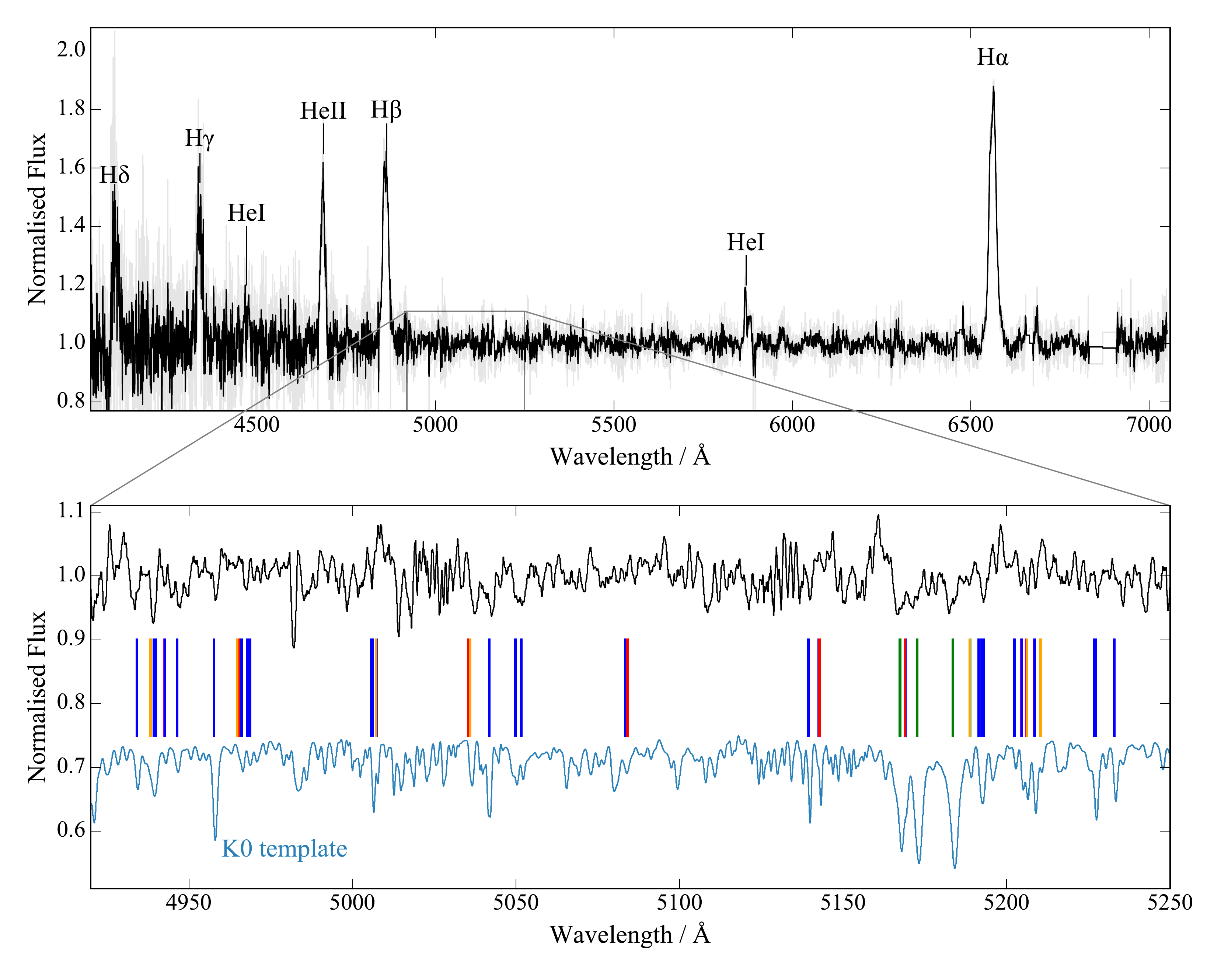}
\end{center}
\caption{Average spectrum for J0644+3344 during the inferior conjunction of the secondary star. $Top:$ We have added all the available spectra around phase 0.0 and normalized to the continuum. We have used a Savitzky-Golay filter with a 5\AA\@ box to smooth the spectrum for clarity. $Bottom:$ Line identification of the donor features compared to a K0 template. We identify lines of \ion{Ca}{i} ($cyan$), \ion{Cr}{i} ($purple$), \ion{Fe}{i} ($blue$), \ion{Mg}{i} ($green$), \ion{Ni}{i} ($red$) and \ion{Ti}{i} ($orange$). }
\label{fig:average}
\end{figure*}

Time-resolved spectroscopy was performed at the Observatorio Astron\'omico Nacional at San Pedro M\'artir, Mexico, on 2008 January 9--15. These were taken using the Echelle spectrograph and the SITe3 CCD detector at the f/7.5 Cassegrain focus of the 2.1 m telescope. Exposures of 600s were used to obtain a resolution of $\phi\sim$0.025 in orbital phase. The log of spectroscopic observations is presented in Table~\ref{tab:spectralog}. We covered a full orbital period on most nights except on January 12, 14 \& 15, when we focused the observations around the primary eclipse. A Th-Ar calibration lamp was used every ten object images in average. The range in spectral coverage is $\lambda\lambda$3980 -- 7050 \AA\@ with an average spectral resolution of $\simeq 20 $ \kms. Standard \caps{iraf}\footnote{IRAF is distributed by the National Optical Astronomy Observatories, which are operated by the Association of Universities for Research in Astronomy, Inc., under cooperative agreement with the National Science Foundation.} procedures were used to reduce the data. No flux standard was observed since we were focused on obtaining radial velocities from the data. Heliocentric corrections have been applied. In Figure~\ref{fig:average}, we present the spectrum of J0644 which shows strong \ion{He}{i}, \ion{He}{ii} and hydrogen Balmer lines in emission. \textbf{Occasionally, we observe narrow absorption lines similar to those presented by S07.}

 \begin{table}
  \caption{Log of Spectroscopic Observations}
  \label{tab:spectralog}
  \centering
 \begin{tabular}{cccc}
   \hline\hline
       Date  & HJD begin &HJD end& No. of \\
    (UT)&+2454000&+2454000&images\\
    \hline
    Jan 09  2008& 474.652536 & 474.953376 & 38 \\
    Jan 10 2008& 475.650633 & 475.944017 & 35 \\
    Jan 11 2008& 476.596902 & 476.997508 & 36 \\
    Jan 12 2008& 477.648375 & 477.838742 & 23 \\
    Jan 13 2008& 478.622799 & 478.904737 & 34 \\
    Jan 14 2008& 479.627510 & 479.711502 & 12 \\
    Jan 15 2008& 480.640736 & 480.766464 & 13 \\
\hline
\end{tabular}
\end{table}

\subsection{Photometry}

Differential photometry was also carried out at the Observatorio Astron\'omico Nacional at San Pedro M\'artir, Mexico, in two seasons. All of the observations were done with the Thomson 2k CCD detector with a 3x3 binning, at the 1.5 m Cassegrain telescope. During the first season, 2008 January 8-17, images were taken using 10s exposures with the $V$ filter. The second one was performed during 2010 November 30 -- December 2. We used the same instrumental setup as before for the first night, and unfiltered images (white light, $WL$) with 3 s exposures for the last two nights. Standard data reduction was done with \caps{iraf}'s \textit{apphot} routines. The images were corrected for bias and flat-field before aperture photometry was carried out. We calibrated the photometry with a star in the field, labelled as J0644-I by S07 with coordinates $RA=~06^h~44^\prime~30.0^{\prime\prime}$ and DEC=~+$33^{\circ}~45^\prime~51^{\prime\prime}$. The log of observations is presented in Table \ref{tab:photolog}, where we have included observations by S07, used in our analysis (see Section \ref{sec:data}). We will refer to each observation through the associated orbital cycle, E, of the primary eclipse.\\

\begin{table*}
\caption{Log of Photometric Observations}
\label{tab:photolog}
\centering
\begin{tabular}{c c c c c c c c}     
\hline\hline
Date  & HJD begin &HJD end    & No. of& HJD&Cycle$^a$ &O -- C&Filter\\
   ( UT )       &+2450000   &+2450000&   images &     mid-eclipse                          &(E)&(s)&\\
    \hline
    
    Feb 02 2005$^b$&3403.63099&3403.86902&300&3403.76076		&0		&105.86&$R$\\
    Feb 04 2005$^b$&3405.61762&3405.87962&319&3405.64518		&7		&2.15&$R$\\
    Mar 01 2005$^b$&3430.60706&3430.76033&107&3430.69582		&100		&-100.23&$R$\\
    Mar 02 2005$^b$&3431.59527&3431.79932&111&3431.77535		&104		&75.85&$R$\\
    \hline
    Jan 09 2008 & 4474.61785 & 4474.95418 & 1599&4474.7925	& 3976	&13.64& $V$\\
    Jan 10 2008& 4475.60821 & 4475.95576 & 1541&4475.8702 	&3980	&25.32&$V$\\
    Jan 11 2008& 4476.60094 & 4477.00459 & 1571& 4476.6764	&3983	&-132.82&$V$\\
    			&			&		&		& 4476.9482			&3984	&69.82&$V$\\
    Jan 12 2008& 4477.59675 & 4477.90855 & 1896& 4477.7597	&3987	&129.84&$V$\\
    Jan 13 2008& 4478.62364 & 4478.93554 & 1912& 4478.8345	&3991	&-132.8&$V$\\
    Jan 14 2008& 4479.60603 & 4479.72369 & 766& 4479.6398		&3994	&-115.65&$V$\\
    Jan 15 2008& 4480.60568 & 4480.79876 & 937& 4480.7177		&3998	&-73.73&$V$\\
   \hline
    Nov 30 2010 &5530.92231	&5531.06772	&  423& 5531.0106	&7897	&-94.75&$V$ \\       
    Dec 01 2010 & 5531.89866	&5531.99292	& 765 & $ \cdots $	&7900	&$ \cdots $&WL\\
    Dec 02 2010 & 5532.82211	&5532.93647	& 776 & 5532.8954	&7904	&8.17&WL\\       
    \hline

\end{tabular}\\
\medskip $^a$Cycles E were calculated with the ephemeris presented in this paper.\\
 $^b$Data by \cite{sea07}.
\end{table*}

\section{Photometric Analysis}
\label{sec:data}

\subsection{TSA and Period04 calculations}

\begin{figure}
\begin{center}
\includegraphics[width=\columnwidth]{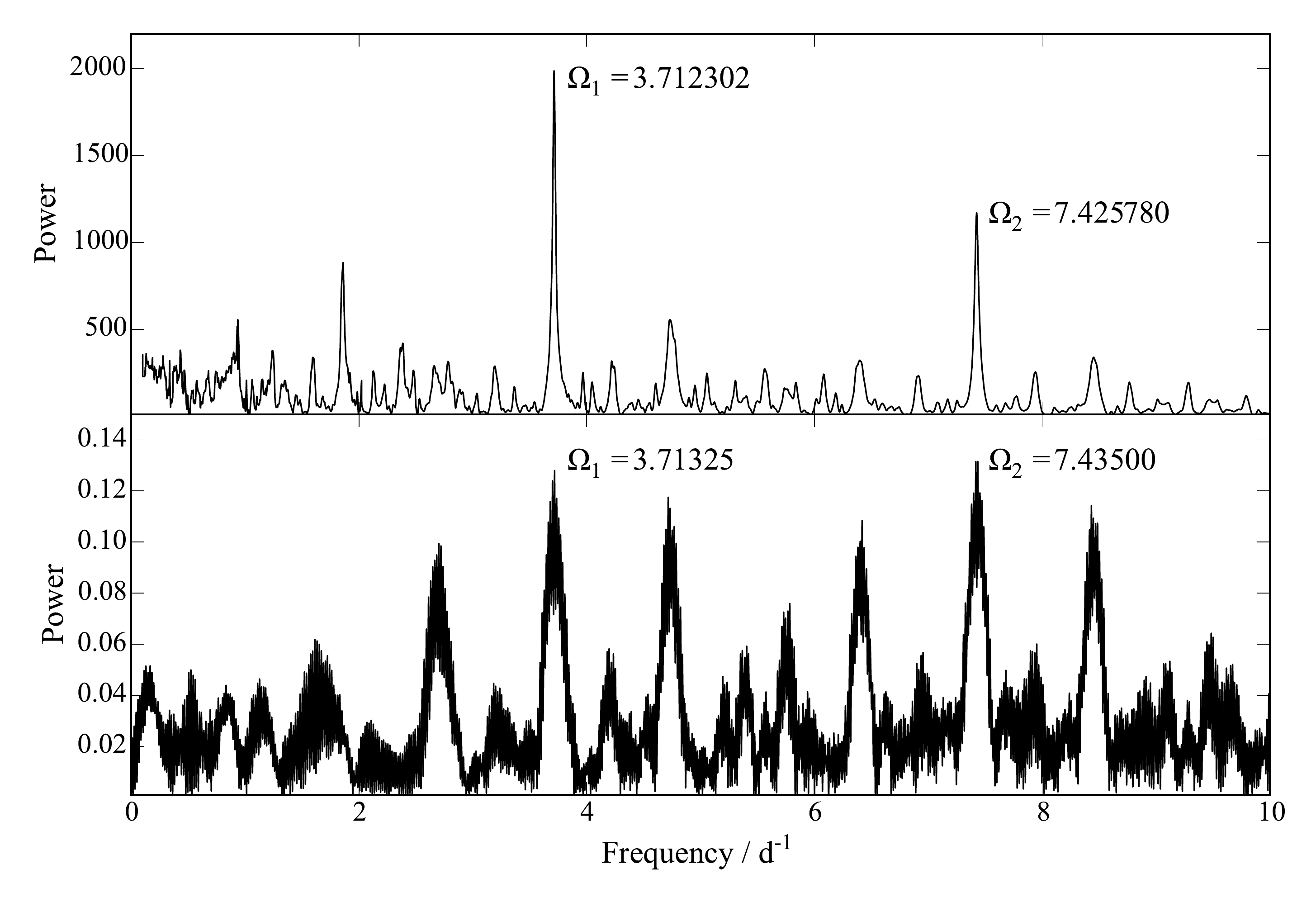}
\caption{Time-series analysis periodograms for J0644+3344 photometry. \textit{Top}. Analysis of Variance ($AOV$). \textit{Bottom}. Discrete Fourier transform. The frequencies orbital period ($\Omega_1$) and its fist harmonic ($\Omega_2$) are labelled.}
\label{fig:periodogram}
\end{center}
\end{figure}

Our photometric data of J0644 were analyzed for periodicities using the analysis of variance (\textit{AOV}) \citep{sch89,dev05} included in the {\scshape vartools} suite \citep{har08} and a discrete Fourier transform with {\scshape period04} \citep{len05}. Both methods show a clear peak around 3.712302 d$^{-1}$ and 3.713254 d$^{-1}$ respectively, shown in Figure \ref{fig:periodogram}. These values correspond to the orbital period of $P\sim6.46$ hr in good agreement with that reported by S07. Other peaks are present in the frequency analysis, but they appear to be main harmonics of the fundamental frequency $\Omega_1$.

\begin{figure*}
\begin{center}
\includegraphics[trim=0cm 0.5cm 0cm 0cm, clip, width=16cm]{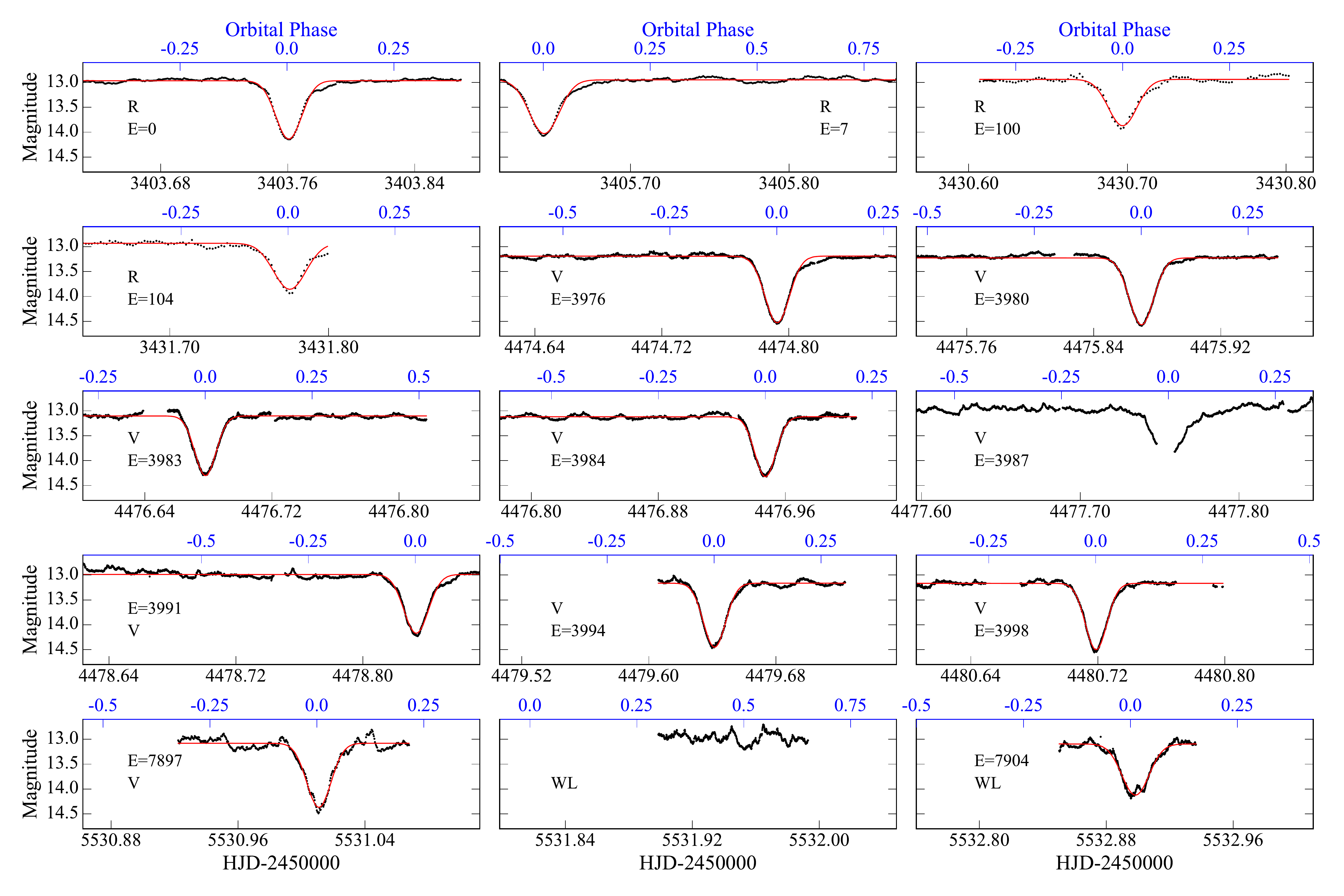}
\end{center}
\caption{Differential Photometry for J0644+3344. All axis are on the same time-scale for comparison. Top axis (in blue) show the orbital phase centred at the corresponding epoch. The filter used on every light curve is indicated; R and V for Johnson $R$ and $V$; WL for white light. Gaussian fits to each eclipse are marked in continuous line (\textit{red} in the electronic version). See text for details of each observation season.}
\label{fig:photall}
\end{figure*}

\subsection{Primary eclipse and improved ephemeris}
\label{sec:ephemeris}

We fitted a Gaussian profile to every fully resolved eclipse and determined the time at mid--eclipse (see Figure~\ref{fig:photall}). These calculated times are shown in Table \ref{tab:photolog}. A linear least--square fit was then applied to the data using the orbital period and zero--phase point from S07 as initial parameters. Preliminary results showed that our ephemeris and those of S07 differ by 0.5 in phase. This was simply a confusion about the time of the inferior conjunction of the secondary. We consulted with the previous authors about this problem and they kindly provide us with their data of February and March 2005 observations (D. Sing, private communication). Their data are included in Table \ref{tab:photolog}. We used their photometry in the $R$ filter, which was also fitted using a Gaussian profile to obtain their mid--eclipse points. These eclipse timings were consistent with our preliminary ephemeris; thus we have combined the data to obtain a larger time--based ephemeris:  $$HJD_{\textrm{mid eclipse}} = 2453403.759533 + 0.26937446E,$$ where phase $\phi=0.0$ is the inferior conjunction of the secondary star of the eclipsed source.

\smallskip

The errors are: $\pm1.8\times10^{-6}$d for $T_0$ and $\pm 1.3\times10^{-7}$ d for the period. We see no evidence of a systemic variation in the O-C values, presented in Table~\ref{tab:photolog}. We should point out that our last eclipse (E7904) has such a complex shape that we did not take it into account in our calculations.

\subsection{The 2008 photometry}
\label{sec:photom08}

\begin{figure*}
\begin{center}
\includegraphics[trim=-0.0cm 0.0cm -.0cm -0.0cm, clip, width=10 cm,angle=270]{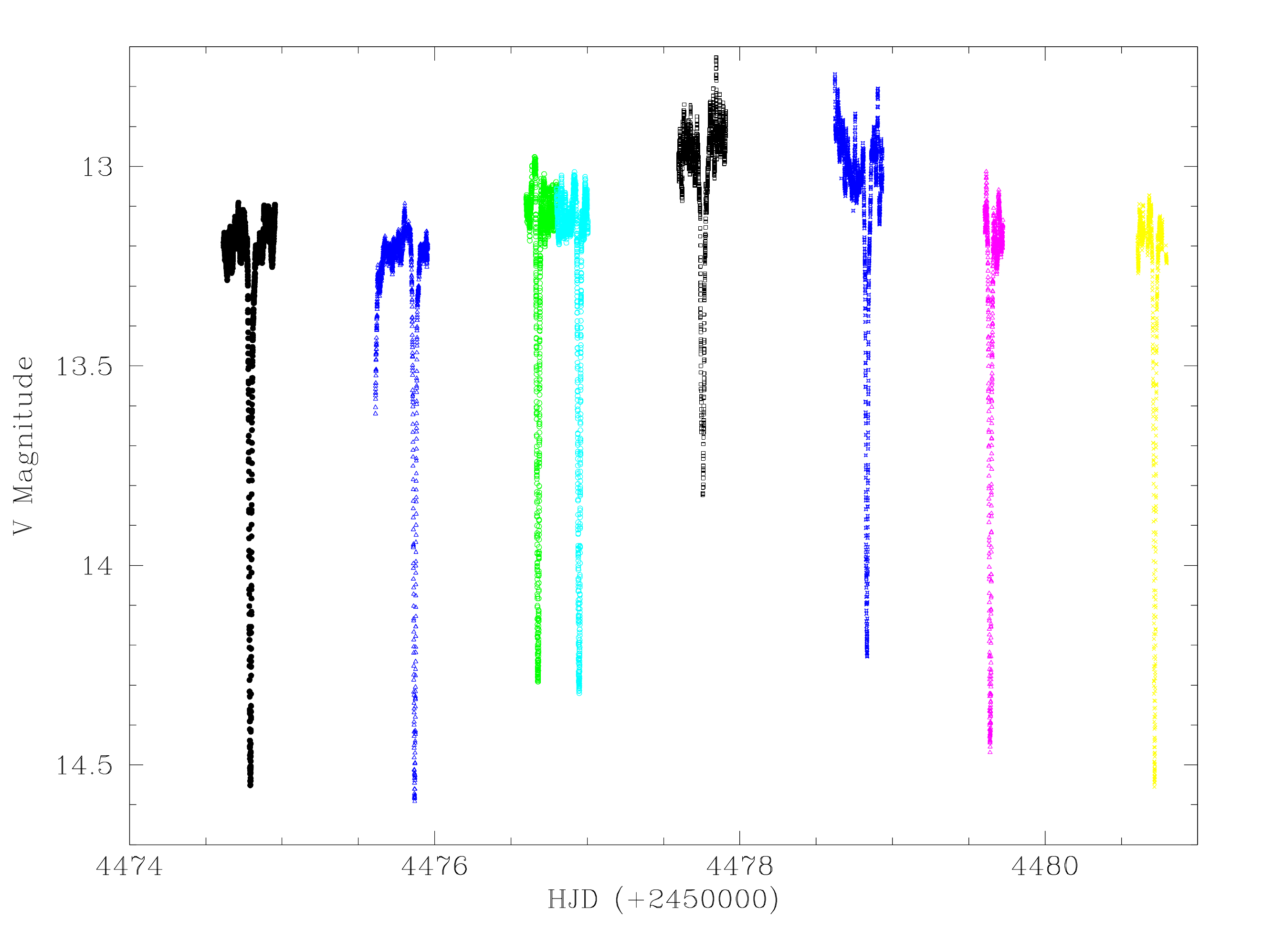}
\end{center}
\caption{Johnson $V$ photometry obtained in 2008 on seven consecutive nights. See text for details.}
\label{fig:phot2008}
\end{figure*}

Our main photometry results originate from the seven consecutive nights taken in 2008 shown in Figure \ref{fig:phot2008}. We observed a small and gradual brightening of $\sim 0.4$ mag towards the fourth night (12.8 mag) and its subsequent decline to the initial 13.2 mag. The depth of the eclipse vary, with a maximum depth of 1.4 mag on the second night to a minimum of 1.2 mag on the fifth night. The two eclipses taken on the third night have almost the same depth. The eclipse on the fourth night was not observed fully, as the observations were interrupted around minimum light (see also Figure \ref{fig:photall}, panel E3987). 

\subsection{Folded photometry}
\label{sec:phot-folded}

\begin{figure*}
\begin{center}
\includegraphics[trim=0.0cm 0.0cm 0.0cm 0.0cm, clip, width=10 cm,angle=270]{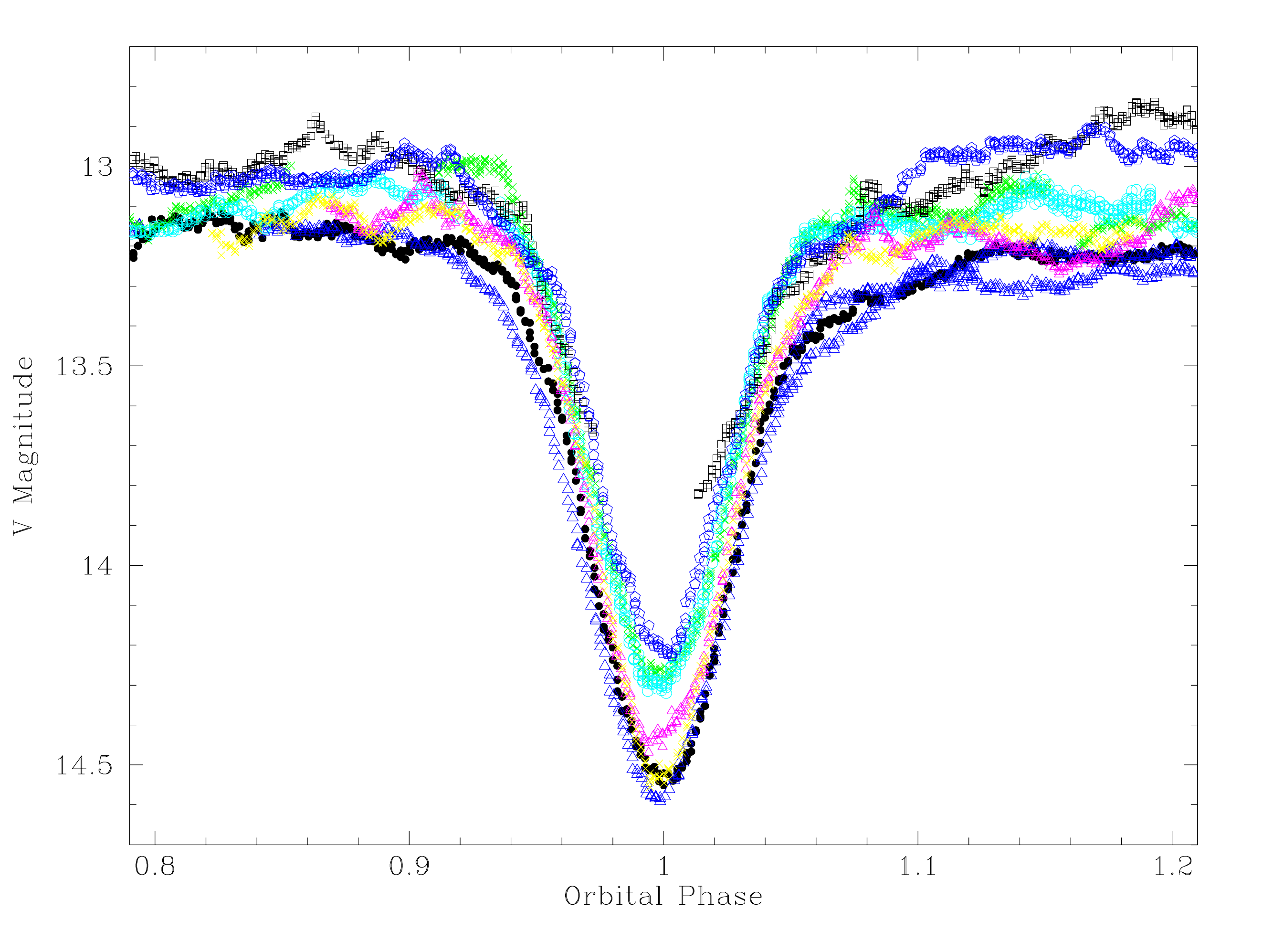}
\end{center}
\caption{Detailed structure of the primary eclipses in 2008. The colours are the same as in Figure \ref{fig:phot2008} in the {\it e-version} }
\label{fig:eclips08}
\end{figure*}

\begin{figure*}
\begin{center}
\includegraphics[trim=0.0cm 0.0cm 0.0cm 0.0cm, clip, width=10 cm,angle=270]{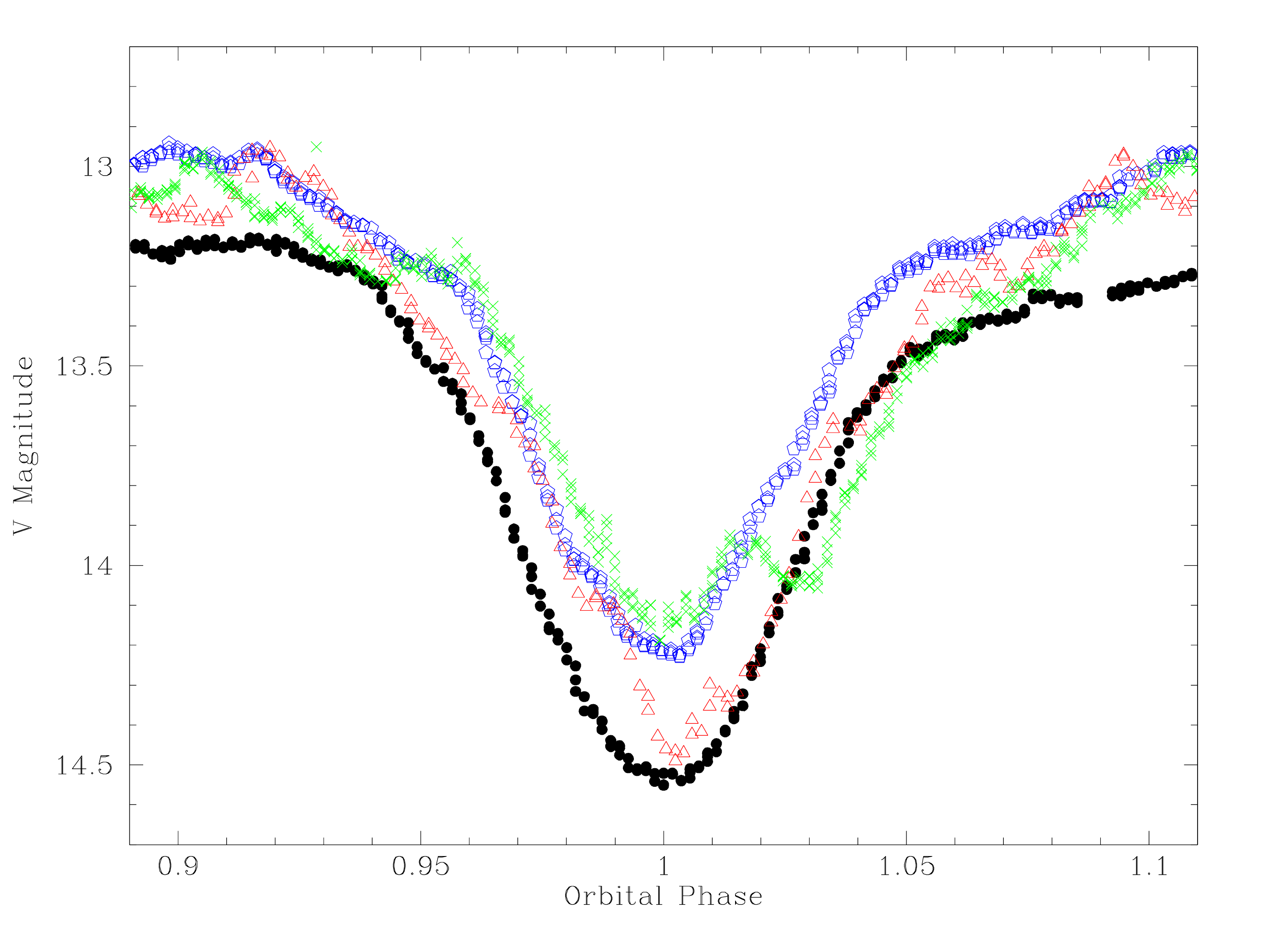}
\end{center}
\caption{Enhanced view of four primary eclipses in 2008 and 2010.  Symbols are as follows: E3991(blue pentagons); E7897 (red triangles); E3976 (black dots); E7904 (green crosses). (Colours seen in the e-version). }
\label{fig:eclipses2}
\end{figure*}

In Figure \ref{fig:eclips08}, we show the 2008 photometric data folded in phase using our ephemeris. In general, the light curves show deep primary eclipses as in the case of the $R$ and $B$ photometry by S07. Although some primary eclipses show a smooth drop in flux with a symmetric rise to the original flux level, others show a more complex shape. The eclipse shape can change from one cycle to the next one, revealing a variable structure inside the primary eclipse. In these light curves, the variability is probably due to changes in the disc distribution brightness. An example of a non symmetrical eclipse is night five (E3991; blue pentagons) where, around phase 0.99, we see small oscillations before minimum light. This was one of the nights when the object was brightest, with $V=13.0$~mag outside eclipse. 
  
This is not the only observed eclipse with an internal structure.  In Figure \ref{fig:eclipses2}, we show an enhanced view of three asymmetrical eclipses plus one with no internal features. These are: E3991 (blue pentagons); E7897 (red triangles) and the WL eclipse E7904 (green crosses). E3976 (black dots) is plotted as the reference of an almost featureless eclipse. The magnitude of these light curves outside eclipse are in the range 13.0--13.2 mag. Several features should be noted in this graph. Between phases 0.91 and  0.97, the three asymmetric light curves show different behaviour, with a smoother decreasing shoulder. They also show a small hiatus around phase 0.98. E7897 shows a marked delay to minimum light reaching 14.5 mag at phase 1.0 and close to the minimum value of E3976. On the egress E7897 shows also a clear feature at phase 1.04. The WL eclipse, E7904 shows a very complex structure. We observe a similar hiatus as E3991 and E7897 at phase 0.98, and then remain close to E3991 up to phase 1.02, after which it shows a second eclipse (shallower than the primary eclipse) with a minimum at phase 1.03. Thereafter, the brightness increases to reach a similar out of eclipse magnitude as E3991 and E7897. In general, these features could possibly be identified with: (a) the outside ingress of the WD+inner region (phase 0.94); (b) the inside ingress of the WD+inner region (phase 0.98); (c) the outside ingress of a hot spot (phase 0.99); (d) the inside ingress of a hot spot (phase 1.1); and (e) the outside egress of the WD+inner region (phases 1.04-1.05). A close inspection of some of the S07 photometry show also evidence of some of these features, even though they were taken with a smaller time resolution. Finally, E3991, E7897, E3983 and E3984 show a slanted bottom, indicating that at times, some of the features in the accretion disk are totally occulted. This implies that the inclination angle is higher than the S07 estimate. This inclination can not be as high as 80$^\circ$, since the evidence reviewed above indicates at most a grazing eclipse of the secondary star on the central emission surrounding the white dwarf. Furthermore, at such inclination angle, we should see a definite secondary eclipse. For these reasons, we will use a conservative value of $i = 78 \pm 2^\circ$, which includes the value used by S07 and includes the upper limit of $80^\circ$.

\section{Radial velocity analysis}
\label{sec:radvel}

\begin{figure}
  \begin{center}
    \includegraphics[width=1.0\columnwidth]{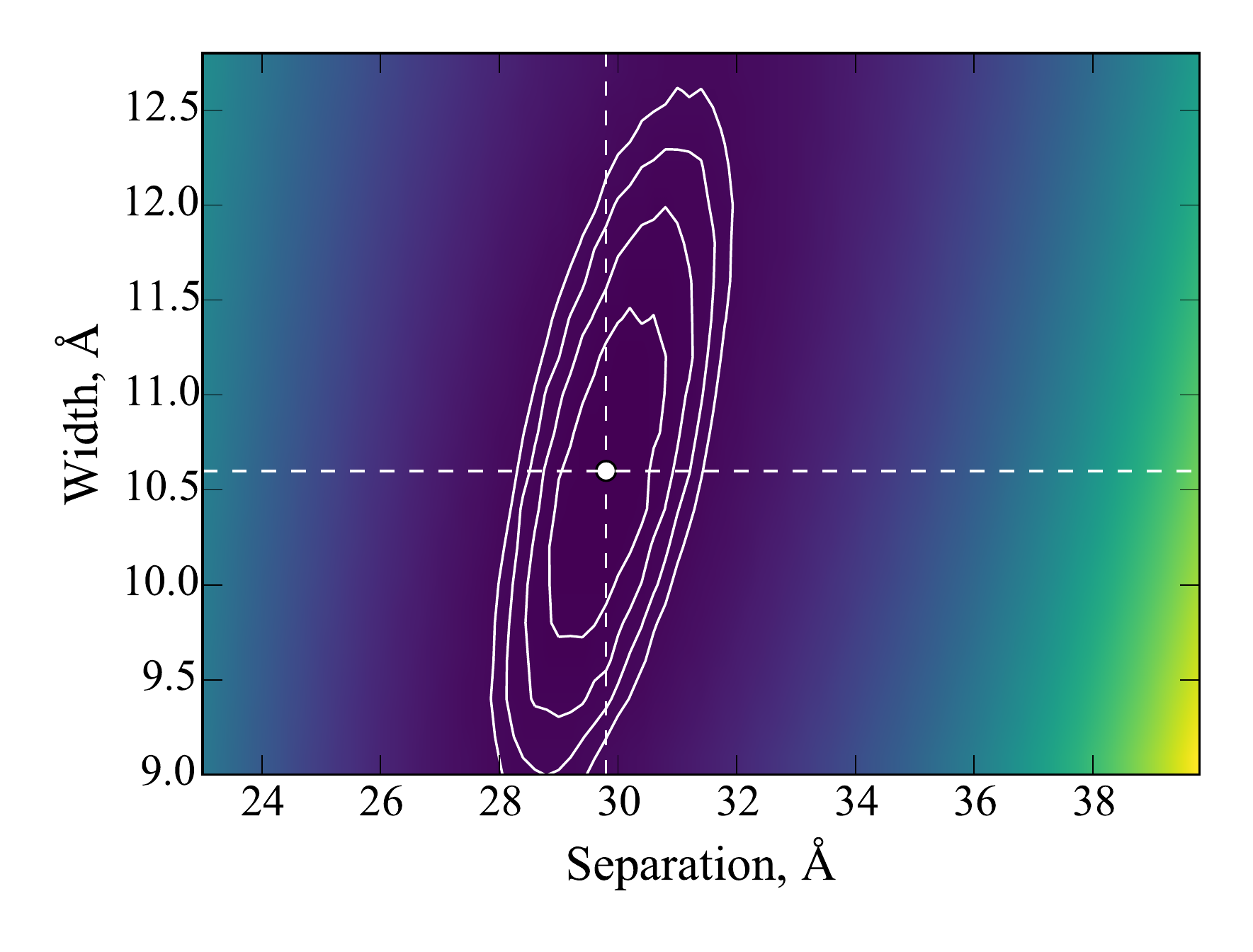}
\caption{Two dimensional map of the control parameter $\Delta \sigma K_1^{-1}$, used to select the best double Gaussian $\alpha=29.8$ \AA\@ and $\omega=10.6$ \AA. Contours represent $\Delta \sigma K_1^{-1}=0.001$.}
    \label{fig:2-d}
  \end{center}
\end{figure}

\begin{figure}
\includegraphics[trim=0.1cm .3cm .4cm 0.2cm,clip,width=8cm]{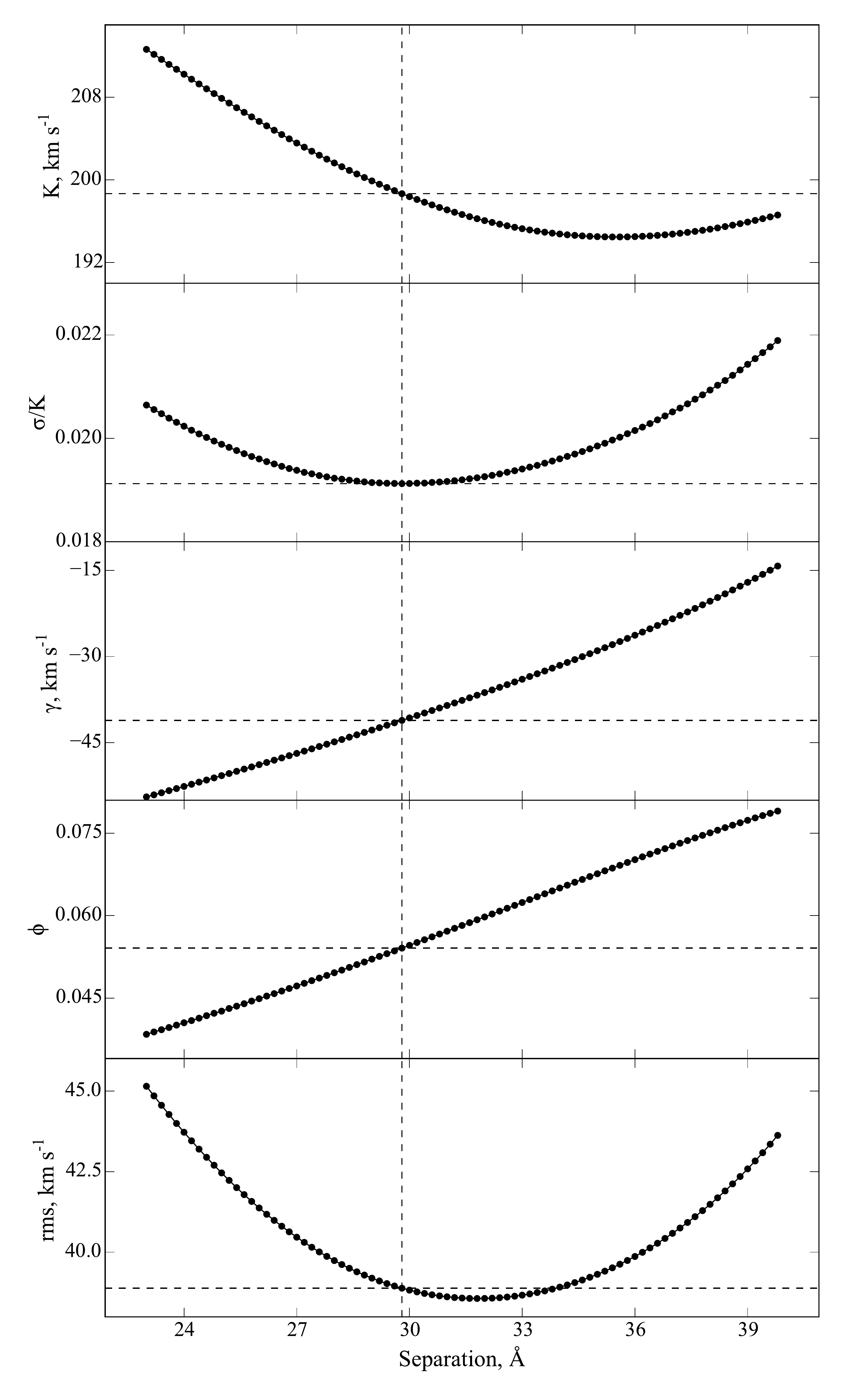}
\caption{Diagnostic Diagram for the Balmer H$_\alpha$ emission line. The parameters are shown as function of the separation of the individual Gaussians. We used $\sigma(K)$ K$^{-1}$ as the best-solution parameter, which corresponds to Gaussian separation of $\alpha$ = 29.8 \AA.}
\label{fig:diagnosha}
\end{figure}

Radial velocity measurements were performed for the hydrogen Balmer lines H$\alpha$ and \ion{He}{ii} $\lambda$4686 \AA\@. For the Balmer line, we used the standard double--Gaussian method with a diagnostic diagram developed by \citet{sha86}, and a single Gaussian fit for the Helium line. This method traces the emission of the wings of the line, presumably arising from the inner parts of the accretion disc, hence, will follow better the motion of the WD. We adapted the IRAF \textit{convrv} routine of the \textit{rvsao} package modified by J. Thorstensen (2008, private communication) to perform a radial velocity fit to a circular orbit:

\begin{equation}\label{eq:orbit}
V \left( \phi \right)=\gamma + K_{em} \sin\left(\phi - \Delta\phi\right)
\end{equation}

where $\gamma$ is the systemic velocity, $K_{em}$ is the semi-amplitude of the radial velocity curve and $\Delta\phi$ is the difference between the photometric and spectroscopic zero-point. We performed a grid search to find the optimal values for the Gaussian separation, $\alpha=$23-40 \AA\@ and a width of $\omega=9-13$ \AA, both in 0.1 \AA\@ steps. We evaluated each individual combination and used $\sigma(K)$ K$^{-1}$ as the standard control parameter to find the minimum, e.g. H$\alpha$ shown in Fig.~\ref{fig:2-d}. We show the results of the analysis for H$\alpha$ in the diagnostic diagram in Figure~\ref{fig:diagnosha}. The $\sigma(K)$ K$^{-1}$ indicator shows a clear minimum value for $\alpha$=29.9 \AA.

Once the optimal radial velocities were found, we created 1000 bootstrap copies of the radial velocity curve and performed the fitting routine to obtain a robust uncertainty on the orbital parameters \citep{efr79}. The distribution of parameter solutions are well described by Gaussians which allowed us to retrieve the 1-$\sigma$ error on each parameter.

\begin{figure}
\centering
\includegraphics[trim=0cm 0cm 0cm 0cm,clip,width=8cm]{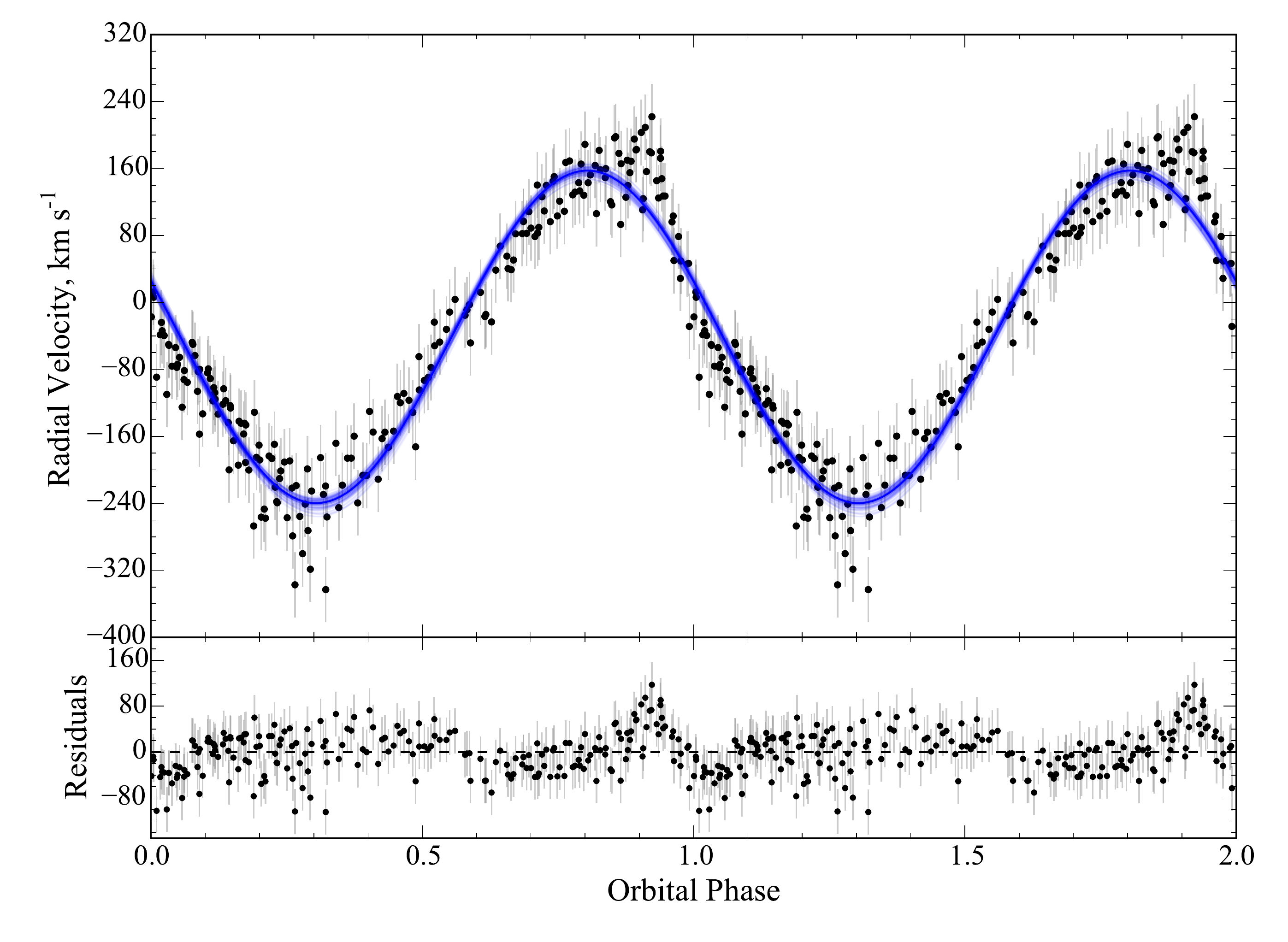}\\
\includegraphics[trim=0cm 0cm 0cm 0cm,clip,width=8cm]{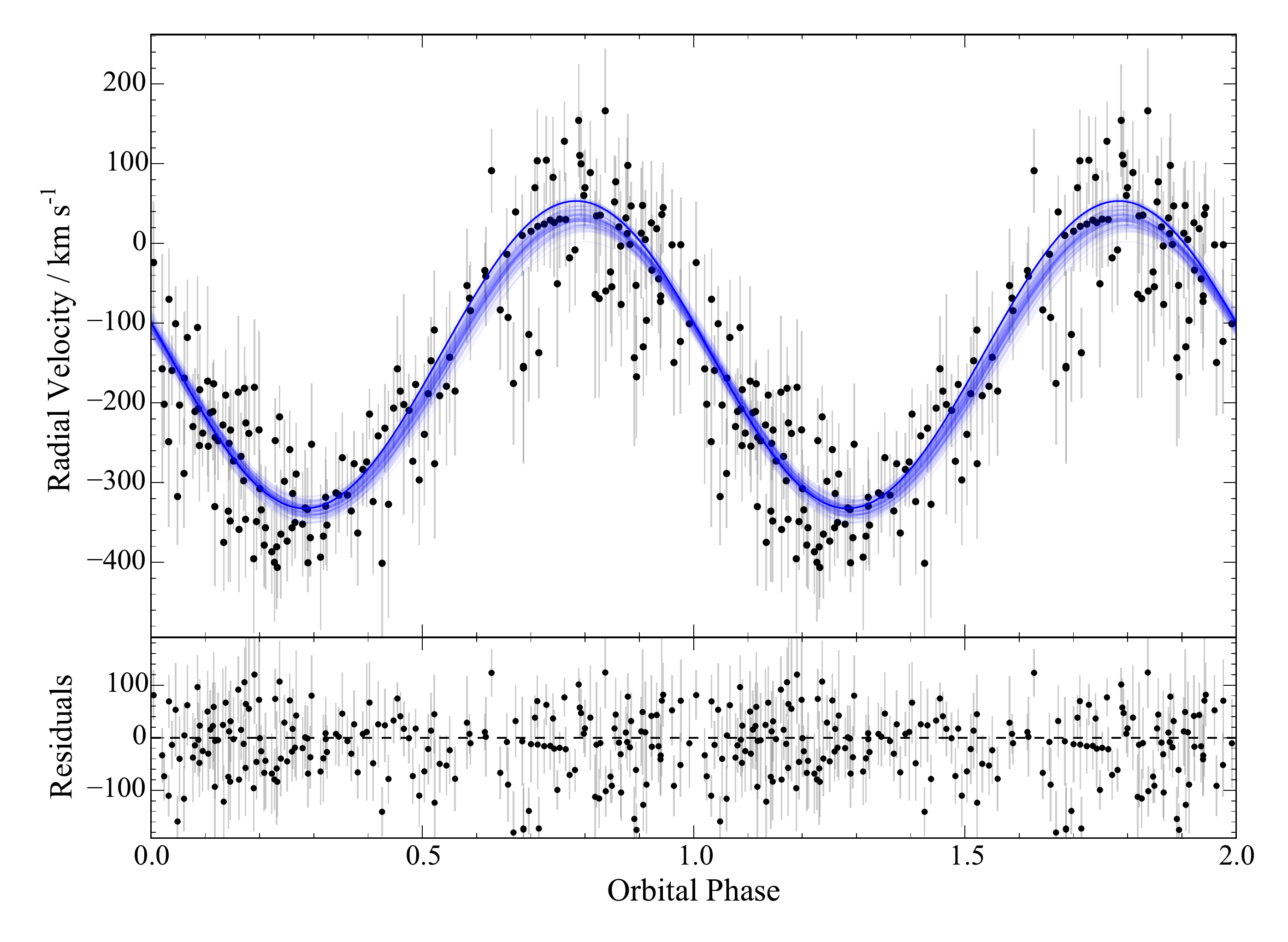}
\caption{Radial velocity curves for H$\alpha$ (\textit{top}) and \ion{He}{ii} $\lambda$4686 \AA\@ (\textit{bottom}) emission lines. The best solution as well random realizations via bootstrapping are shown (\textit{blue lines}) to reflect the scatter of the solution. Errors on individual data points have been scaled so $\chi^2_{\nu}=1$. }
\label{fig:prim_vel}
\end{figure}

The best solution for the radial velocities is shown in the top panel of Fig.~\ref{fig:prim_vel}. The solid curve correspond to the fits using the values presented in Table \ref{tab:radial_vel}, where we show the parameter values obtained with the least--square routine and the errors from the bootstrap technique. These errors have been scaled so $\chi^2_{\nu}=1$. The optimal parameters are presented in Table~\ref{tab:radial_vel}. 

The \ion{He}{ii} line is single-peaked throughout the full orbit. We performed a single Gaussian fit to every spectra and obtain the corresponding radial velocities. We fitted Eq.~\ref{eq:orbit} to the data and obtained the best parameters, shown in Table~\ref{tab:radial_vel}. Errors on the parameters were obtained via 1000 bootstrap copies in a similar way as for H$\alpha$. The radial velocity curve is shown in the bottom panel of Fig.~\ref{fig:prim_vel}. These results are discussed in Section \ref{sec:discussion}, where we compare the double--Gaussian method with the method used by S07.

\begin{table}
\caption{Radial velocity solutions for the emission lines in J0644.}
\begin{center}
\begin{tabular}{lccc}
\hline
\label{tab:radial_vel}
Line& $K_1$& $\gamma$&$\phi - \phi_{phot}$\\
& \kms\ & \kms &\\
\hline
H$\alpha$			& 198 $\pm$ 4	&-41  $\pm$ 3& 0.05  $\pm$ 0.01 \\
\ion{He}{ii} $\lambda$4686 	& 184 $\pm$ 8	&-150 $\pm$ 6&-0.04  $\pm$ 0.01 \\
\hline
\end{tabular}
\end{center}
\end{table}

\section{The donor star}\label{sec:donor}

S07 found the strongest evidence of the donor by performing cross--correlations with several standard stars with spectral types K3--K5. Although we see several weak absorption lines, most of these features are masked by the bright emission arising from the accretion disc during most of the orbit. Thus, we were unable the find a consistent solution to the radial velocities. We created an average spectra around phase zero, where the absorption lines are most likely to observed (see Figure~\ref{fig:average}). The spectrum is consistent with a K0V star. The usual lines for spectral classification like \ion{Fe}{i}$\lambda \lambda$ 4250, 4260, 4271 and \ion{Cr}{i} lines $\lambda \lambda$ 4254, 4274, 4290 as well as \ion{Ca}{i} $\lambda$ 4226, are too weak to be seen in this heavily masked spectrum due to the presence of the strong blue continuum. We have therefore resorted to a direct comparison with the K0V star $\sigma$ Draconis and compared the region 4900-5250, wich is shown in the blow-out segment at the bottom of the Figure. It is well known  that secondary stars in CVs tend to have larger radius than main sequence stars for a given mass \citep{ech83,bea98,kni06}. Therefore, obtaining its mass from the spectral type or a mass-radius relation is not straightforward \citep[the mass of a K0V is $\sim0.89$ \msun,][]{fea98}. Using empirically calibrated relations, we can estimate a donor mass M$_2 = 0.85\pm0.22$ \msun  \citep{kni06,kni11}. However, this relationship is uncertain above 0.6 \msun, hence we favour the dynamical measurement of K$_2$ obtained by S07 in our mass estimate in Section \ref{sec:discussion}. Nevertheless, both methods give similar results, within the errors.

\section{Doppler tomography}
\label{sec:doppler}

\begin{figure*}
\centering
\includegraphics[trim=0.3cm 0.7cm 0.3cm 0.6cm,clip,width=7.5cm]{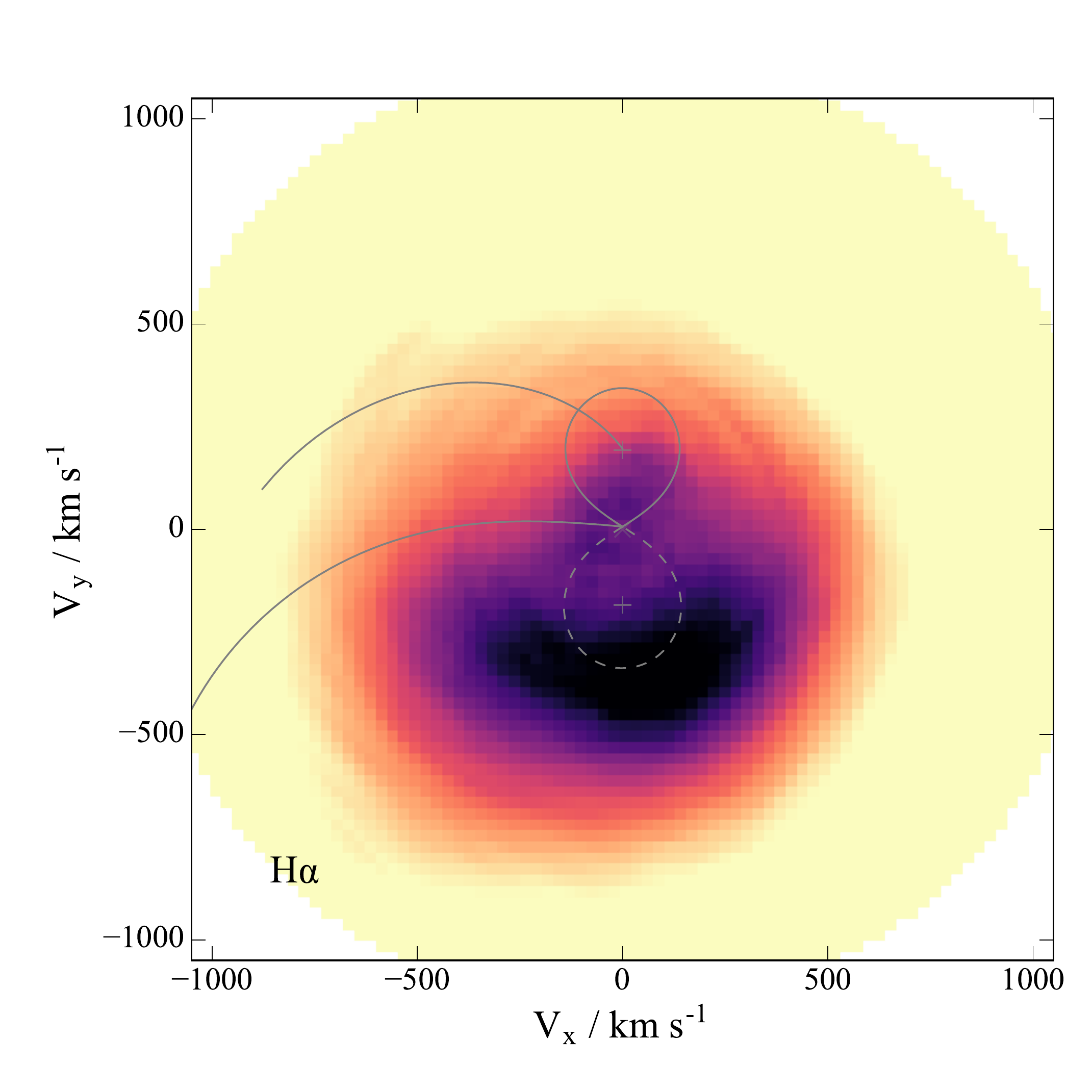}
\includegraphics[trim=0.3cm 0.7cm 0.3cm 0.6cm,clip,width=7.5cm]{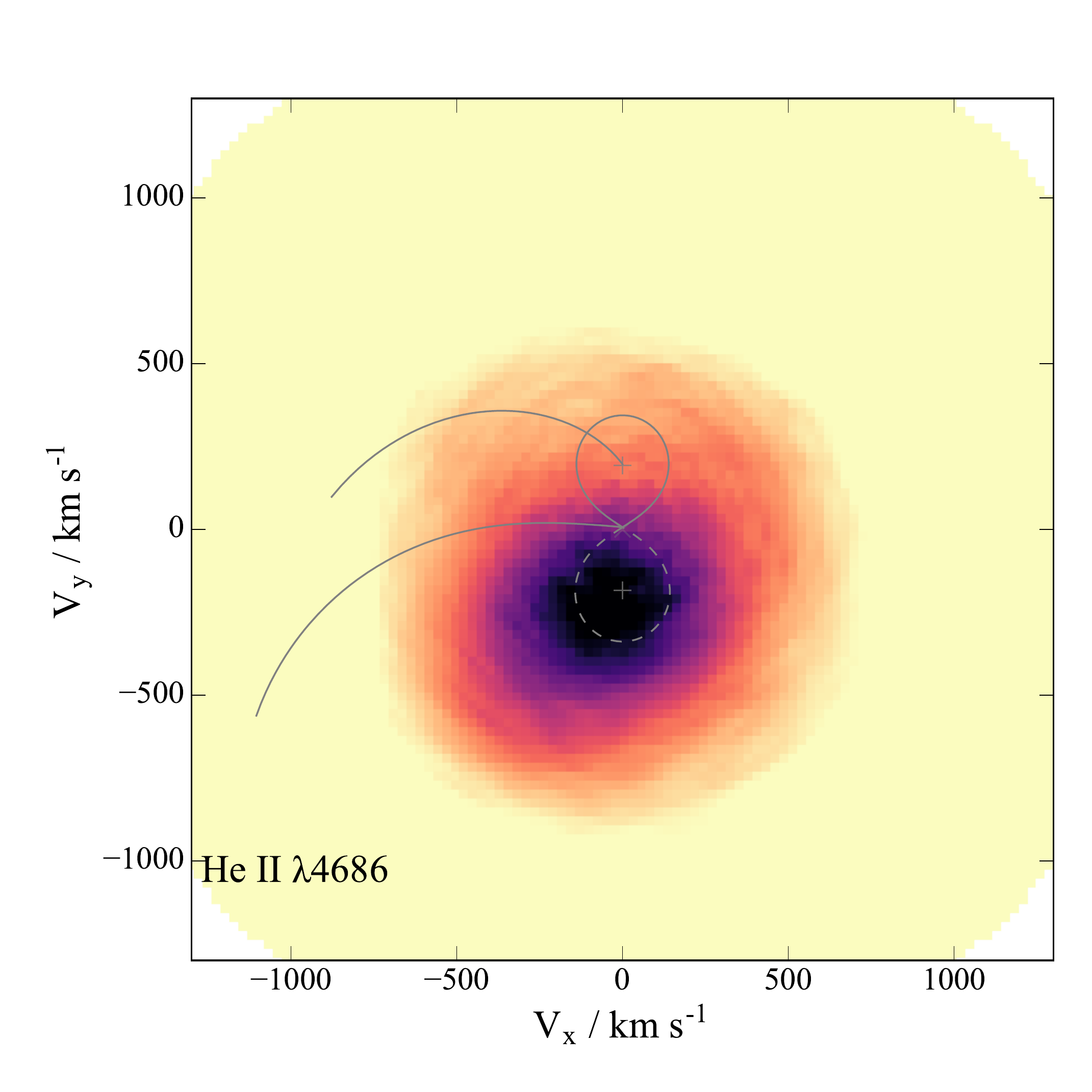}\\
\includegraphics[trim=0.3cm 0.7cm 0.3cm 0.6cm,clip,width=7.5cm]{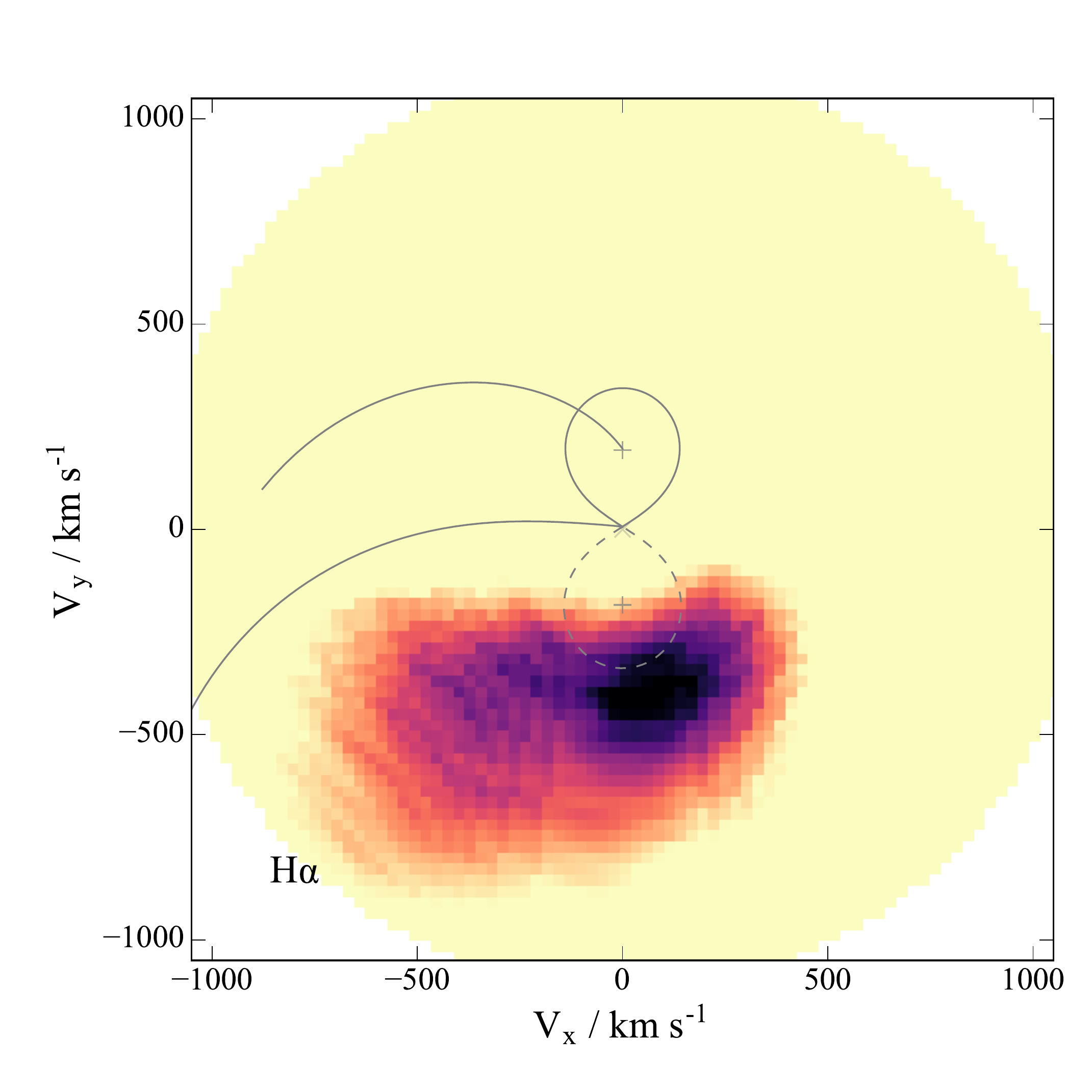}
\includegraphics[trim=0.3cm 0.7cm 0.3cm 0.6cm,clip,width=7.5cm]{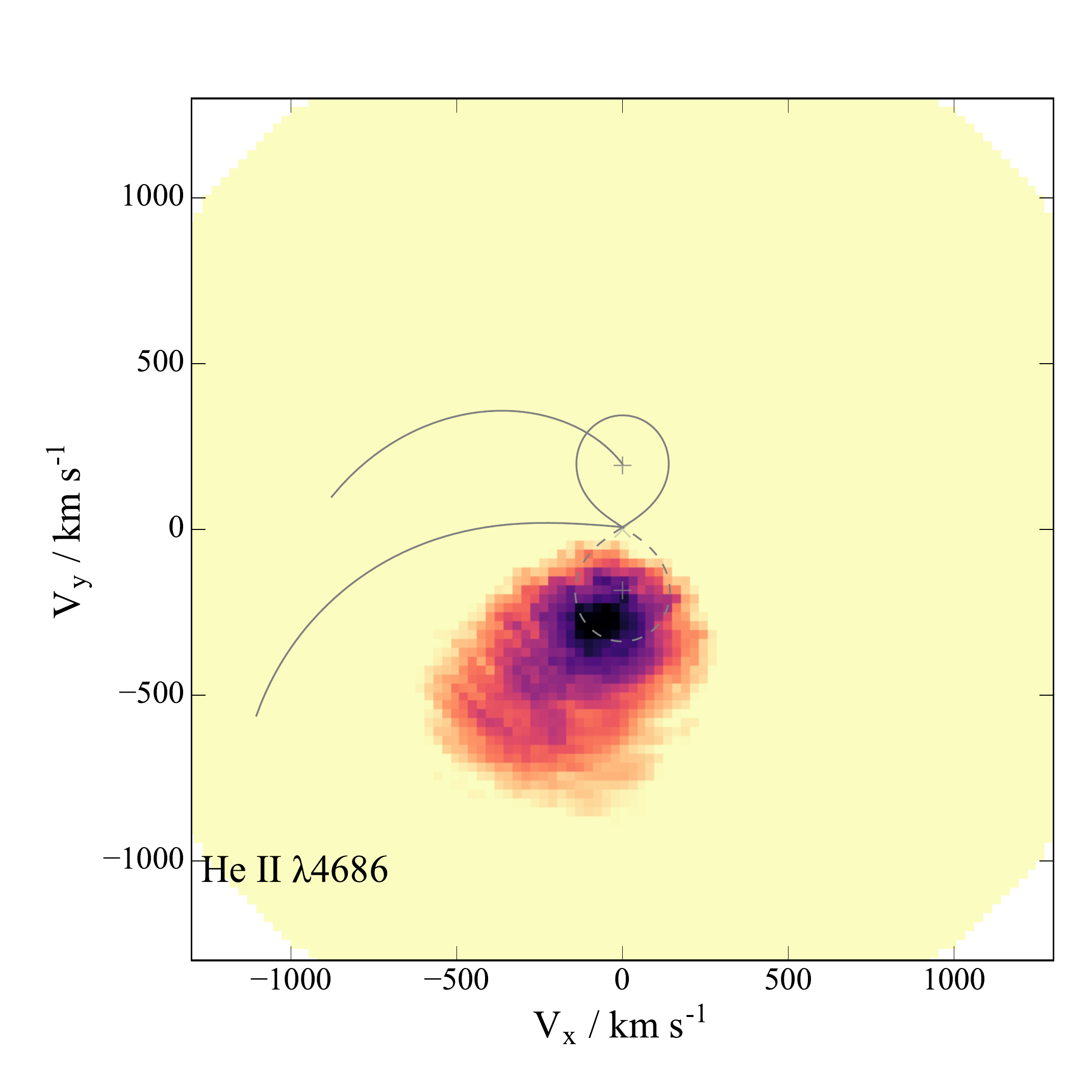}
\caption{Doppler Tomography of J0644 for H$_\alpha$ ($left$) and \ion{He}{ii} $\lambda$4686 \AA\@ ($right$) emission lines. The Roche lobe surface was calculated using the orbital parameters discussed in this paper: $i=78^\circ$, $q=0.96$ and $\gamma=-7.1$ \kms. The Keplerian and ballistic trajectories in the upper figures are marked as the upper and lower curves, respectively. The crosses are the velocity (from top to bottom) of the secondary star, the center of mass and the primary star. (See text for discussion). The bottom figures show the same tomograms after subtracting an azimuthal average to highlight the asymmetric emission components. }
\label{fig:doppler}
\end{figure*}

Doppler tomography provides a method to obtain insight on the structure of the accretion disc, seen through a particular emission and/or absorption line \citep{mah88}. They are created in velocity space and careful interpretation is needed to infer the physical position from where they are emitted. We used a code developed by \citet{spr98} based on the maximum entropy method\footnote{\url{http://www.mpa-garching.mpg.de/~henk/pub/dopmap/}}. We calibrated the spectra using the $V$ Johnson simultaneous photometry to account for slit losses. An average of the simultaneous photometric data, was performed, covering exactly each spectroscopic exposure and then summed all the spectral fluxes centred at 5380 \AA\@ with a width of 98 \AA. Then, we obtained a $V$ Johnson-like flux (not convolved with the filter spectral response). The spectra were then calibrated accounting for the difference between photometric and spectroscopy magnitudes and correcting accordingly. Doppler maps were thus obtained for the emission lines H$\alpha$ and \ion{He}{ii} 4686 \AA\@ (Figure \ref{fig:doppler}).

The Balmer H$\alpha$ tomogram shows an emission region with a maximum at ($V_x$,$V_y$)$\simeq$(+200,-300) \kms~ and a bright spot close to the secondary star (near the L1 point). In the higher contrast map (bottom row of Figure~\ref{fig:doppler}), this spot appears connected with the larger region at negative $V_y$ velocities. The H$_\beta$ tomogram (\textit{center panels}) presents a more extended emission region, ranging from ($-V_x$,$-V_y$)=(-200,-700) \kms~ to ($-V_x$,$-V_y$)=(500,0) \kms. The high excitation line \ion{He}{ii} $\lambda 4686$ ( \textit{right panels}) is seen in the tomogram as a single blob of material around or close to the WD centred at ($V_x$,$V_y$)= (+0,-200) \kms. These tomograms are consistent with previous studies on NL-type systems \citep{kai94,ns11}.

At phase $\phi=0.0$, the Balmer lines are visible even tough the system undergoes deep eclipses. A minimum is found at phase 0.125 (consistent with the tomography), and in fact show a double peak throughout phase 0.375 as different parts of the large emission is occulted. In general, the Balmer lines present a complex but similar behaviour during the full cycle. The blue-shifted peak in both lines is more intense throughout phase 0.0 to 0.25, probably due to the strength shown in the tomography at the negative $V_x$ part of the elongated emissions. There is a clear double speak at H$\alpha$ present at phases 0.375 and 0.75, phases at which we would expect to see the hot spot near the secondary more clearly and separated from the blob in its less elongated position. We most stress that the high-excitation \ion{He}{ii} $\lambda$4686~\AA\@ is always present as a single peak at all phases, although weaker near phase zero. The \ion{He}{ii} line shows a minimum at phases 0.0 to 0.125, but does not disappear completely. This suggests that the blob we are looking in the tomogram may surround completely the central WD and may arise from outside the orbital plane, but is could also be the results of seeing the extended weaker emission shown in the lower right tomogram of Figure~\ref{fig:doppler}).

\section{Discussion}
\label{sec:discussion}

Accurate masses of CVs are difficult to obtain, even in eclipsing systems, since it is rarely possible to look directly at the WD. Instead, we rely on the radial velocity curve of the emission lines arising from the accretion disc, which can posses asymmetries that may lead to inaccurate results. Furthermore, the surface of the secondary star could be irradiated and the measurements from its absorption lines could give an out of center of mass radial velocity semi-amplitude. These problems have been discussed in full by \citet{war95}, although modern studies with high resolution spectroscopy have proven useful in de-convolving these asymmetries, as in the case of U Gem \citep{eche07}.

Our radial velocity analysis of J0644 clearly indicate that the disc has large asymmetries, as shown in the Balmer line. This is consistent with the findings by S07, particularly with their results for the \ion{He}{i} $\lambda 4471$ line in absorption (see their Fig. 9). Our Doppler Tomography shows that the Balmer emission comes from a large, but asymmetric region located on the far side of the accretion disc with respect to the secondary. We must point out, however, that our observations do not show the strong central absorption in the \ion{H}{i} and \ion{He}{i} lines observed by S07. We must therefore assume that the accretion disc was at a different state. However, the \ion{He}{ii} tells us another story. While S07 ascribe the bulk of the \ion{He}{ii} emission coming from in inner-disc, our Doppler tomography results indicate that the \ion{He}{ii} emission arise from a concentrated region around the white dwarf. Nevertheless, these two interpretations should not affect the $K_{He II} $ value. The S07 results for the \ion{He}{ii} line gives $K_{He II} = 151 \pm 5$ \kms, compared with our results of $K_{He II} = 184 \pm 8$~\kms.  We point out at a strong difference between our Figure~\ref{fig:prim_vel} and S07 Figure 11. In both cases the positive radial velocities reach a maximum of about 100 $km \, s^{-1}$. However, our maximum negative velocities reach about $-350$~\kms, while in S07 they go no further than $-250$~\kms. S07 measured the centre-of-wavelength, from each line profile to calculate the Doppler velocity of \ion{H}{i}, \ion{He}{i} and \ion{He}{ii} lines. They found, in general that the radial velocity curve from these lines (in absorption and emission) present a flat bottom from phases 0.0 - 0.5, which they attribute to line blending, magnetic effects, non-uniform disc emission or effects from an accretion stream. They also claim that \ion{He}{ii} in emission is exempt from this non-sinusoidal behaviour. This may not be so. Although the solid line in their Figure 11 is a sinusoidal, this may be misleading, as the radial velocity measurements do seem to be affected by this flattening behaviour between phases 0.1 to 0.4. The cause of this effect may be due simply to the way their individual spectra was measured. The centre-of-wavelength approach may result in distorted radial velocities. This method may not be the cause {\it per se}. It is possible that the asymmetric bulk may play a role in this distortion in both \ion{H}{i} and \ion{He}{i} lines, when at these phases, the bulk is behind the white dwarf. But it is also possible that we have observed the system at a different accretion stage and the flattening may be due to a stronger distortion in both \ion{H}{i} and \ion{He}{i} lines. This possibility is supported by the fact that we do not observe the strong central absorption in the \ion{H}{i} line. We must also point out that our \ion{H}{i} radial velocity curve does not suffer from this flattening effect. If anything we observe an excess of negative radial velocities at phase 0.25. If our arguments are correct, then our \ion{He}{ii} radial velocity should reflect more accurately the motion of the white dwarf. This value is also supported by our Doppler Tomography (see section \ref{sec:doppler}), which indicate that the centroid of the velocity \ion{He}{ii} blob has a velocity around $\sim200$~\kms.  

S07 estimates a semi-amplitude of the secondary of $K_2=192.8\pm5.6$ \kms. We will use the latter in our estimates of the orbital parameters. Therefore, we can establish an independent mass ratio of $q=K_1/K_2=0.96\pm0.05$. We adopted our improved orbital period, $P_{\scriptsize\textrm{orb}}=0.26937446$ d to estimate the mass functions:

$$M_1 \sin^3 i = {P K_2 (K_1 + K_2)^2 \over 2 \pi G} = 0.77 \pm 0.06 M_{\odot},$$

$$M_2 \sin^3 i = {P K_1 (K_1 + K_2)^2 \over 2 \pi G} = 0.73 \pm 0.07 M_{\odot},$$
and
$$ a \sin i = {P (K_1 + K_2) \over 2 \pi} = 2.01 \pm 0.05 R_{\odot}.$$

Assuming an inclination angle of $i =78 \pm 2^\circ$, the system parameters are: 
\begin{center}
$M_1 = 0.82 \pm 0.06$ $M_\odot$\\
$M_2 = 0.78 \pm 0.07$ $M_\odot$\\
$a = 2.05 \pm 0.06$ $R_\odot$,\\
\end{center}
where the errors reflect both the masses and inclination angle uncertainties. At these high inclination angles, the error produced by its uncertainty is small and the results are dominated largely by the error in the masses and the separation of the binary. 

Our results yield larger masses for both the primary and secondary stars (within the uncertainties) than the upper limits given by S07. The separation of the binary is also slightly larger than that estimated by S07. However, these purely dynamical measurements provide a mass of the WD in line with the average mass of CVs \citep{zor11}. The mass of the donor is in agreement with the expected value from semi-empirical evolutionary sequences \citep{kni11}, as calculated in Section~\ref{sec:donor}. 

\subsection{The narrow absorption spectra}
\label{sec:narrow}

Although we detected the narrow absorption features mainly in the Balmer lines, around phase 0.5, we were unable to reproduce the results of S07, due do the lack of the appropriate spectra. S07 used two spectra taken at phases -0.0788 and -0.0177 and subtracted them with the purpose to find if the spectrum of the WD reveals itself. What they found is a spectrum (see their Figure 11) with strong narrow absorption lines of H I, \ion{He}{i}, \ion{Mg}{ii} and \ion{Ca}{i}, instead of the broad features expected from a white dwarf. They concluded that the region producing this spectrum could arise from a region containing the WD and an inner-accretion disc, or could be the result of some newly formed pre-WD. Our Doppler tomography of H$\alpha$ and \ion{He}{ii} $\lambda$4686 \AA, suggest that this absorption spectrum is the result of the accretion stream overflow (i.e. the disc regions detected through the Balmer lines) seeing at large optical depths against the bright inner regions around the WD (e.g. \citet{hr94}). This supports the explanation by S07 that we are looking at a region with deep \ion{H}{i} and \ion{He}{i} absorption lines with a black body temperature of about 25,000 K.

\subsection{An SW Sextantis system?}
\label{sec:membership}

J0644 exhibits most of the characteristics of an SW Sex type star, as defined originally by \citet{tho91,hoa03}. It is a nova-like CV showing deep primary eclipses and a high inclination system. It displays high excitation lines including \ion{He}{ii} $\lambda$ 4686 \AA. It shows single peak emission lines, instead of the double-peaked emissions expected from near-edge-on discs. It shows shallower Balmer lines during primary eclipse and also transient absorption lines around phase 0.5. Two characteristics of the SW Sex stars do not comply for J0644. It has an orbital period way outside the 3--4 hr criteria and it does not exhibit pronounced radial velocity phase offsets with respect to phase zero, as defined by the eclipses. There are now several new found CV's with large orbital periods like BT~Mon, V363~Aur, AC~Cnc and LS~Peg definitely confirmed as SW ~Sex systems, plus a number of possible candidates. Therefore, this last criteria might be only a selection effect, as it should be the criteria of having only eclipsing system in order to belong to this group. Our Doppler tomography shows that the Balmer lines are produced in J0644 by elongated and strong emissions, located opposite the secondary star with respect to the center of mass of the binary. This is similar to the Doppler tomography found in BH~Lyn \citep{dea92}. A small phase shift could be produce if the main emitted material is located opposite the secondary star. Another case that shows a concentrated emission at different parts of the disc is the case of BF~Eri \citep{nz08} shows also a small phase lag.

\section{Conclusions}
\label{sec:conclusions}

We have improved the ephemeris of the object through differential photometry. A slow and short brightening of $\sim$~0.4~mag was observed in 2008. Our radial velocity analysis allowed us to obtain values for the semi-amplitudes of both components of the system. Doppler tomography revealed the \ion{He}{ii} emission arises from the WD and is a good indicator of its radial velocity semi-amplitude. From our adopted values for $K_1$, $K_2$ and $i=78\pm 2^{\circ}$, we find $M_1 = 0.82 \pm 0.06$~M$_\odot$; $M_2 = 0.78 \pm 0.04$~M$_\odot$ and a separation of the binary $a = 2.05 \pm 0.06$~R$_\odot$. We found that the general characteristics of J0644 are consistent with a SW Sex nova-like CV. Follow--up observations, specially simultaneous spectroscopy and photometry, are needed to better understand the nature of this object.

\section*{Acknowledgments}

The authors are indebted from DGAPA (Universidad Nacional Aut\'onoma de M\'exico) support, PAPIIT projects IN111713 and IN122409. JVHS acknowledges financial support from CONACyT (Mexico) scholarship programs, the MSc. program at IA-UNAM and the University of Southampton. The authors would like to thank all the staff at the OAN for their invaluable help. This research has made use of NASA's Astrophysics Data System. We would like to thank the anonymous referee for the prompt response and excellent feedback.

\bsp

\label{lastpage}

\end{document}